\documentclass[12pt]{iopart}
\usepackage{iopams}  
\usepackage{graphicx}
\def\be{\begin{equation}}
\def\ee{\end{equation}}

\newcommand{\beq}{\begin{equation}}
\newcommand{\eeq}{\end{equation}}
\newcommand{\ber}{\begin{eqnarray}}
\newcommand{\eer}{\end{eqnarray}}
\newcommand{\berr}{\begin{eqnarray*}}
\newcommand{\eerr}{\end{eqnarray*}}

\begin{document}

\title{ 
Zero mode contribution in quarkonium correlators and in-medium
properties of heavy quarks
}

\author{Saumen Datta$^a$ and P\'eter Petreczky$^b$ 
}
\address{ 
$^a$  Department of Theoretical Physics, Tata Institute of Fundamental
Research, Homi Bhabha Road, Mumbai 400005, India\\
$^b$ Department of Physics and RIKEN-BNL Research Center, Brookhaven
National Laboratory, Upton, New York, 11973
}

\begin{abstract}
We calculate the low energy contribution to quarkonium correlators in Euclidean time 
in lattice QCD. This contribution was found to give  the dominant source of the
temperature dependence of the correlators. We have found that the low
energy contribution is well described by a quasi-particle model and have 
determined the effective temperature dependent heavy quark mass.
\end{abstract}

\pacs{11.15.Ha, 11.10.Wx, 12.38.Mh, 25.75.Nq}



%
\section{Introduction}
\label{intro}

Quarkonium correlators in Euclidean time and the corresponding spectral functions at finite temperature have been
extensively studied in the past few years in connection with the problem of quarkonium melting in the
deconfined phase \cite{datta04,umeda02,asakawa04,jako07}. Somewhat surprisingly these studies have indicated that
quarkonium states can survive up to temperatures as high as $1.6T_c$, contradicting 
the expectations based on potential models with screening (see e.g. \cite{digal01,mocsy06,mocsy07}). 
To get definite answer to the question at which temperature quarkonium states melt and possibly resolve this
contradiction a very detailed understanding of the temperature dependence of quarkonium correlators is needed.

Above the deconfinement temperature quarkonium spectral functions contain information not only 
about quark anti-quark pairs (whether bound or un-bound) but also about scattering states of single heavy quarks in the plasma. 
The later shows up as a contribution to the spectral function at
very low frequency, $\omega \simeq 0$.  
Therefore this contribution is referred to as the zero mode contribution. It has been realized that the zero
mode contribution could be the dominant source of the temperature dependence of the Euclidean 
correlators in the vector channel \cite{derek} as well as in the scalar and 
axial-vector channels \cite{umeda07} (see also the discussion in Refs. \cite{mocsy07,alberico}). 
In this paper we are going to
study in detail the zero mode contribution and its relation to in-medium properties of heavy quarks.  

\begin{figure}
\includegraphics[width=8cm]{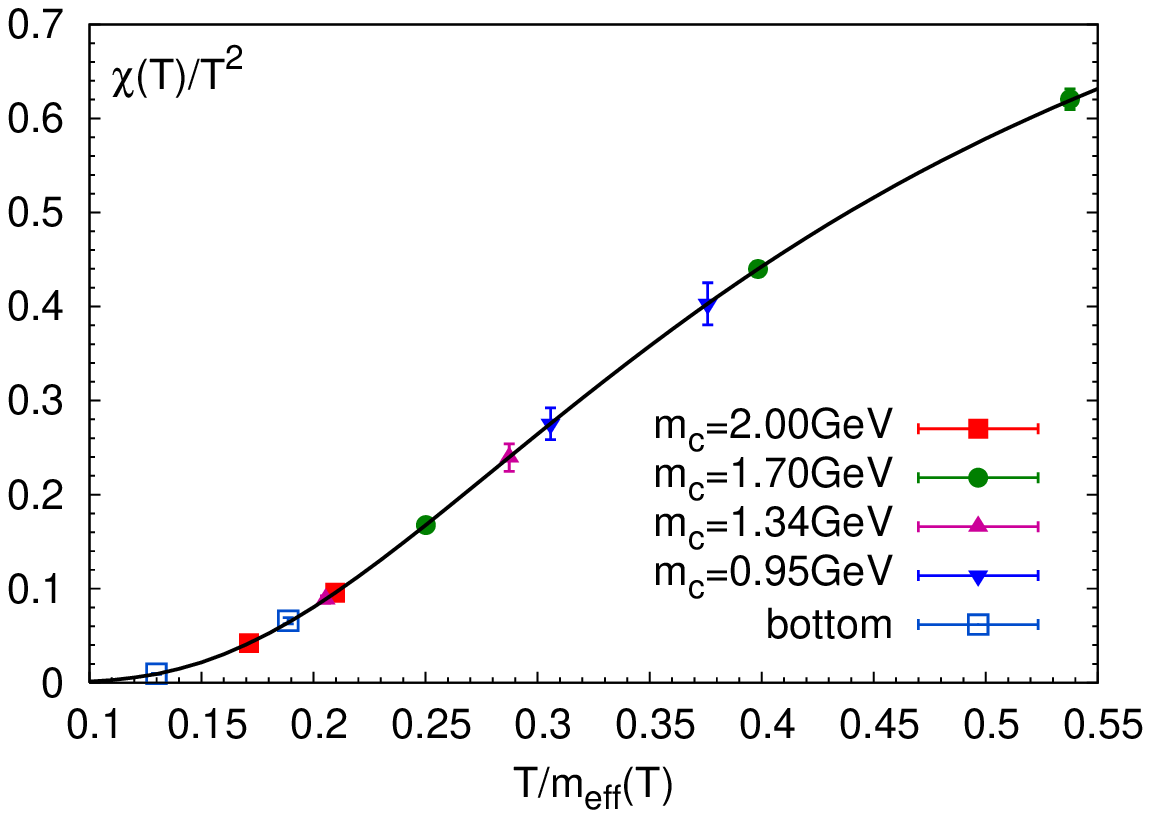}
\includegraphics[width=8cm]{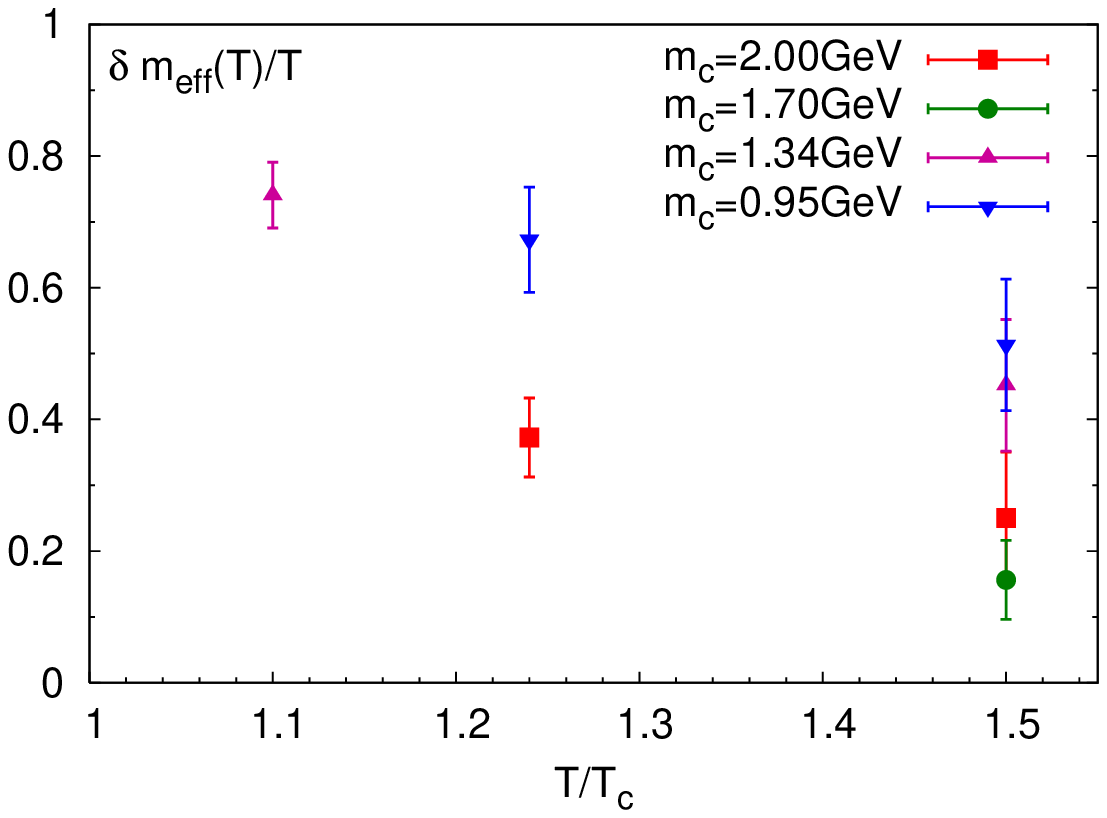}
\vspace*{-0.4cm}
\caption{Quark number susceptibility for heavy quarks (left) and the thermal correction to
the effective quark mass (right). The line shows the prediction of the quasi-particle model with $m_{\rm eff}(T)$.}
\end{figure}

\section{Numerical Results}
We have calculated correlators of local quarkonium operators, $\bar q(x) \Gamma q(x)$, 
in quenched approximation using isotropic lattices with
standard Wilson gauge action and non-perturbative clover action for quarks.  Calculations have been performed 
at three different values of the gauge
coupling $\beta=6/g^2=6.499,~6.640$ and $7.192$ corresponding to lattice 
spacing $a=0.0451{\rm fm},~0.0377{\rm fm}$ and $0.017$ fm. The lattice spacing
has been set using the Sommer scale $r_0=0.5$ fm. 
We consider several quark masses in the region of the charm quark mass as well as bottom quark. 
In order to quantify the temperature dependence we have performed calculations both at low temperatures
(below the deconfinement temperature)  and at high temperatures. At low temperatures the spectral functions 
can be calculated reliably with the Maximum Entropy Method and the algorithm of Ref. \cite{jako07}.
Using this spectral function we can calculate the reconstructed correlator 
\begin{equation}
\vspace*{-2cm} G_{rec}^i(\tau, T) = \int_0^{\infty} d \omega
\sigma^i(\omega,T^*)  \frac{\cosh(\omega(\tau-1/2
T))}{\sinh(\omega/2 T)}.
\label{eq.kernel}
\end{equation}
and study the temperature dependence of the ratio $G^i(\tau,T)/G^i_{rec}(\tau,T)$ \cite{datta04}.
Here $T^*$ is some temperature below $T_c$ and $i$ labels the quantum number channel.
The trivial temperature dependence due to
the integration kernel cancels in this ratio. For sufficiently large quark mass ($m \gg T)$ the spectral
function can be written as 
$\sigma^i(\omega,T)=\sigma_{\rm low}^i(\omega,T)+\sigma^i_{\rm high}(\omega,T)$ and similar expression
can be written for the correlator (see discussion in Ref. \cite{derek}). 
The high energy part, $\sigma^i_{\rm high}(\omega,T)$ has contribution
for $\omega>2 m$ and describes the propagation of quark anti-quark pair. 
In the free theory $\sigma^i_{\rm low}(\omega,T)=\chi^i(T) \omega \delta(\omega)$, with $\chi^i$ being calculated in Ref. \cite{aarts}. 
In the presence of interaction the delta function will become a 
Lorentzian with a width $\sim T^2/m \ll T$ and therefore the low energy contribution
$G^i_{\rm low}(\tau,T)\simeq T \chi^i(T)$ is independent of $\tau$ to very good approximation. 
It turns out that this contribution provides the dominant source
of the temperature dependence of the quarkonium correlators. 
To see this we have calculated the time derivative of the correlator $G^{i'}(\tau,T)$, where this (almost) constant 
contribution drops out and studied the temperature dependence of the ratio $G^{i'}(\tau,T)/G^{i'}_{rec}(\tau,T)$. 
We find that this ratio is close to one, more precisely
it deviates from unity by at most $5\%$ 
and always compatible with one within statistical errors up to temperatures as high as $3T_c$! One may think that this
implies the survival of  $\chi_c$ states up to these temperatures. However, as has been shown in 
Ref. \cite{mocsy07} the melting of the P-states does not lead
to significant change in the correlators. and the ratio $G^{i'}(\tau,T)/G^{i'}_{rec}(\tau,T)$ always stays close to one. 
\begin{figure}
\includegraphics[width=8cm]{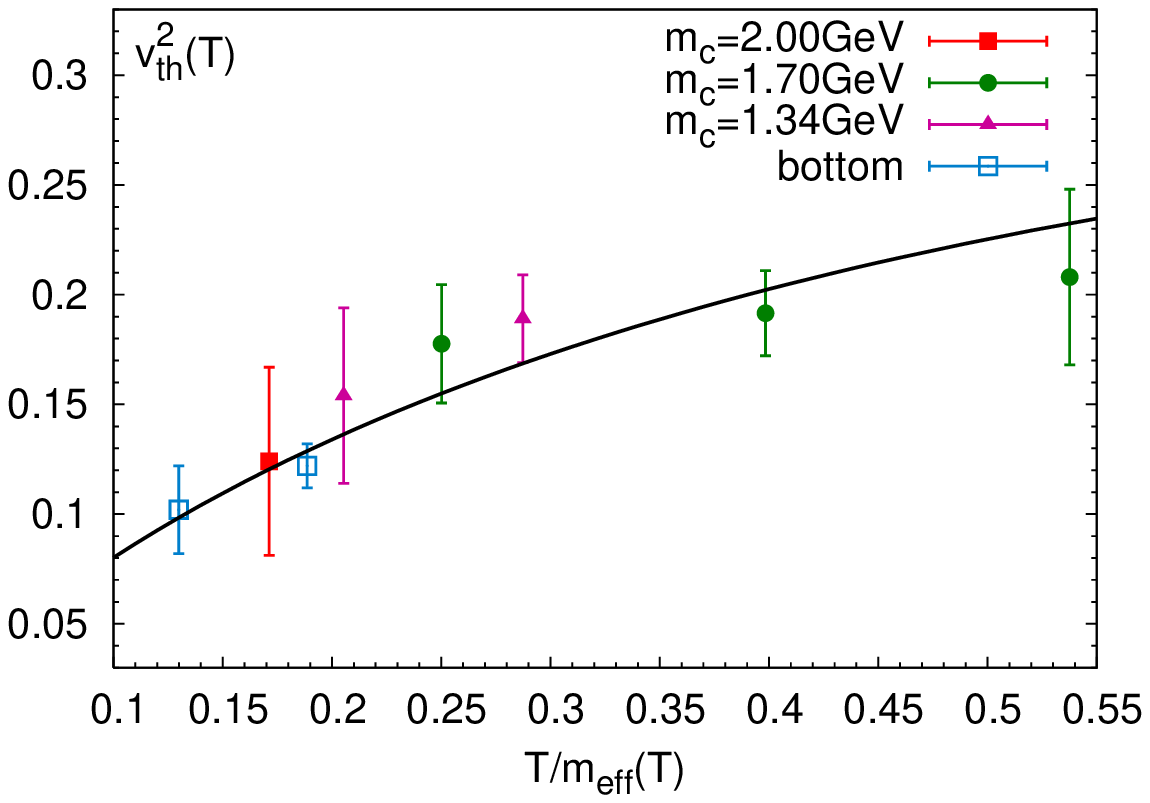}
\includegraphics[width=8cm]{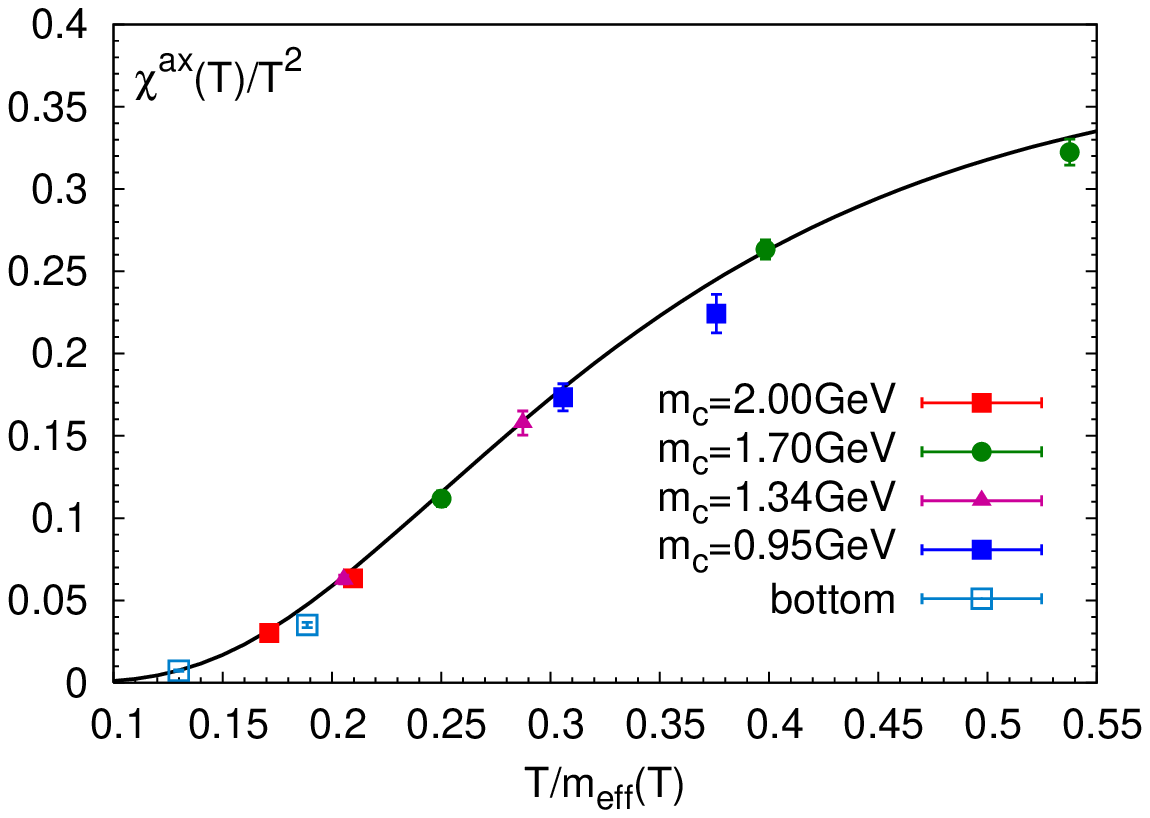}
\vspace*{-0.4cm}
\caption{Thermal velocity of heavy quarks (left) and the constant contribution to the axial-vector correlator (right). The lines show
the prediction of the quasi-particle model with $m_{\rm eff}(T)$.}
\end{figure}

Since we expect that almost the entire temperature dependence of the 
correlator is contained in $G_{\rm low}$ we can assume that $G^i_{\rm high} = G^i_{rec}$ and
quantify the low energy part $G^i_{\rm low}$ above $T_c$ using the lattice data. 
First consider the temporal component of the vector correlator $G^{V0}$. In this
case there is no high energy component 
and $G^{V0}=T \chi^{V0}(T) \equiv T \chi(T)$, with $\chi(T)$ being the heavy quark number susceptibility. In the free theory
$\chi/T^2$ depends only on $m/T$. 
Thus matching the free theory expression for $\chi^{V0}=\chi(T)$, calculated in Ref. \cite{aarts}, to the lattice data on $G^{V0}$, we can define
and effective temperature dependent heavy quark mass $m_{\rm eff}(T)$.  
In Fig. 1 we show the quark number susceptibility as function of $T/m_{\rm eff}$ as well
as the effective quark mass. 
As one can see from the figure the quark number susceptibility is function of $T/m_{\rm eff}$ only. 
The thermal correction to the heavy quark mass is largest at $T_c$ and monotonically decreases with increasing temperature. 
For the charm quark mass of $1.7GeV$ it is negligible for $T>1.5T_c$.
Next we consider the the spatial component of the vector correlator. 
We find that the quasi-particle picture with the effective heavy quark mass determined above
describes the data very well. 
In Fig. 2 we show the ratio $G_{\rm low}^{Vs}(T)/G^{V0}$ which gives an estimate of the thermal velocity squared 
of the heavy quark $v_{\rm th}^2$. 
This is because in the free theory $G_{\rm low}^{Vs}(T)/G^{V0} \simeq \int d^3p (p^2/E_p^2) e^{-E_p/T}/\int d^3 p e^{-E_p/T}=v_{\rm th}^2$.
For $T \gg m$ we have $v_{\rm th}^2 \simeq T/m$ but as one can see from the figure 
it is not a good approximation even for the b-quark.
We also consider the zero mode contribution in the axial-vector channel. The zero mode contribution can be extracted in the same
manner as in the vector case. In Fig. 2 we also show $G^{\rm ax}_{\rm low}/T^2=\chi^{\rm ax}(T)/T^2$ as function of the effective mass
$m_{eff}/T$ together with the prediction of the quasi-particle model. The figure clearly shows that the zero mode contribution
to the axial-vector channel is the function of the effective quark mass only and is in excellent agreement with the
prediction of quasi-particle model.

\section{Conclusions}

In this paper the temperature dependence of quarkonium correlation function has been discussed. It has been shown
that the temperature dependence of the high energy part of the spectral function, for example, the melting of resonances does
not lead to a large change in the correlation function. The dominant source of the temperature dependence of quarkonium
correlators is the zero mode contribution. This contribution has been studied quantitatively on the lattice  for the first
time. In general, it  is expected that this contribution  depends on the temperature and the quark mass. We have found, however,  that it is the function
of $m_{eff}/T$ only and is well described by a quasi-particle model down to temperatures as low as $1.1T_c$.
We have also found that the thermal corrections to the heavy
quark mass are small. Since the the quasi-particle mode is so successful in describing the zero mode contribution to
the quarkonium correlators it would be of great interest to calculate it systematically in improved perturbation theory \cite{scpt}. 

\section*{Acknowledgments}
This work was supported by U.S. Department of Energy under
Contract No. DE-AC02-98CH10886. We thank A. Jakov\'ac for providing his code
for the Maximum Entropy Method analysis.

\vfill\eject

\vskip0.5truecm

  
\vskip0.5truecm
\begin{thebibliography}{99}




\bibitem {datta04}
S.~Datta et al,
Phys.\ Rev.\ D \textbf{69}, 094507 (2004)

\bibitem {umeda02}
T.~Umeda et al,
hep-lat/0211003
 
\bibitem {asakawa04}
M.~Asakawa and T.~Hatsuda,
Phys.\ Rev.\ Lett.\ \textbf{92}, 012001 (2004)

\bibitem{jako07}
  A.~Jakov\'ac et al,
  Phys.\ Rev.\  D {\bf 75}, 014506 (2007)


\bibitem{digal01}
  S.~Digal et al,
  Phys.\ Lett.\  B {\bf 514}, 57 (2001);
  Phys.\ Rev.\  D {\bf 64}, 094015 (2001)

\bibitem{mocsy06}
  A.~M\'ocsy and P.~Petreczky,
  Phys.\ Rev.\  D {\bf 73}, 074007 (2006)


\bibitem{mocsy07}
  A.~M\'ocsy and P.~Petreczky,
  Phys.\ Rev.\ Lett.\  {\bf 99}, 211602 (2007); 
  Phys.\ Rev.\  D {\bf 77}, 014501 (2008)

\bibitem{derek}
  P.~Petreczky and D.~Teaney,
  Phys.\ Rev.\  D {\bf 73}, 014508 (2006)

\bibitem{umeda07}
  T.~Umeda,
  Phys.\ Rev.\  D {\bf 75}, 094502 (2007)

\bibitem{alberico}
  W.~M.~Alberico et al
  Phys.\ Rev.\  D {\bf 77}, 017502 (2008)

\bibitem{aarts}
  G.~Aarts and J.~M.~Martinez Resco,
  Nucl.\ Phys.\  B {\bf 726}, 93 (2005)

\bibitem{scpt}
  F.~Karsch, et al,
  Phys.\ Lett.\  B {\bf 401}, 69 (1997);
  J.~O.~Andersen et al,
  Phys.\ Rev.\ Lett.\  {\bf 83}, 2139 (1999);
  J.~P.~Blaizot et al,
  Phys.\ Rev.\ Lett.\  {\bf 83}, 2906 (1999)





\end{thebibliography}
\end{document}